\documentclass[aps,
prl,
twocolumn,
superscriptaddress,
groupedaddress]{revtex4}  

\usepackage{amssymb,amsmath}
\usepackage{graphicx}
\usepackage{dcolumn}
\usepackage{bm}

\usepackage{hyperref}
\usepackage[usenames, dvipsnames]{color}

\newcommand{\parr}[2]{\frac{\partial{#1}}{\partial{#2}}}

\newcommand{\txt}[1]{\textrm{#1}}

\newcommand{\non}{\nonumber}

\begin{document}

\title{Speed of sound and baryon cumulants in heavy-ion collisions}

\author{Agnieszka Sorensen}
\email{agnieszka.sorensen@gmail.com}
\affiliation{%
 Department of Physics and Astronomy, University of California, Los Angeles, CA 90095, USA
}%
\affiliation{%
Lawrence Berkeley National Laboratory, 1 Cyclotron Road, Berkeley, California 94720, USA
}%

\author{Dmytro Oliinychenko}
\affiliation{%
	Institute for Nuclear Theory, University of Washington, Box 351550, Seattle, Washington 98195, USA
}%

\author{Volker Koch}
\affiliation{%
Lawrence Berkeley National Laboratory, 1 Cyclotron Road, Berkeley, California 94720, USA
}%

\author{Larry McLerran}
\affiliation{%
	Institute for Nuclear Theory, University of Washington, Box 351550, Seattle, Washington 98195, USA
}%

\date{\today}

\begin{abstract}

We present a method that may allow an estimate of the value of the speed of sound as well as its logarithmic derivative with respect to the baryon number density in matter created in heavy-ion collisions. To this end, we utilize well-known observables: cumulants of the baryon number distribution. In analyses aimed at uncovering the phase diagram of strongly interacting matter, cumulants gather considerable attention as their qualitative behavior along the explored range of collision energies is expected to aid in detecting the QCD critical point. We show that the cumulants may also reveal the behavior of the speed of sound in the temperature and baryon chemical potential plane. We demonstrate the applicability of such estimates within two models of nuclear matter, and explore what might be understood from known experimental data.

\end{abstract}

\maketitle


\section{Introduction}
\label{introduction}

The speed of sound, $c_s$, is a fundamental property of any substance. In fluids, it is the velocity of a longitudinal compression wave propagating through the medium, and its square is computed as the ratio of a change in the pressure, $P$, corresponding to a change in the energy density, $\mathcal{E}$. Therefore, it is directly related to the thermodynamic properties of the system, including its equation of state (EOS). 

In dense nuclear matter, $c_s$ is of particular interest to neutron star research: its behavior as a function of baryon number density, $n_B$, influences the mass-radius relationship and, consequently, the maximum possible mass of neutron stars \cite{Ozel:2016oaf}. Current neutron star data suggest that $c_s$ rises significantly for an $n_B$ larger than the nuclear saturation density, $n_0$, and that it perhaps exceeds
$c_s\sim1/\sqrt{3}$ at densities as low as a few times that of normal nuclear matter. This possibility was first suggested in \cite{Bedaque:2014sqa}, followed by other studies, e.g.,\ \cite{Tews:2018kmu,McLerran:2018hbz,Fujimoto:2019hxv}. 

Presently, heavy-ion collisions are the only means of studying dense nuclear matter in a
laboratory. Experiments probing nuclear matter at high $n_B$, such as the Beam
Energy Scan program at the Relativistic Heavy Ion Collider (RHIC), put special significance on the search for the QCD critical point (CP). Here, $c_s$ also conveys relevant information: it displays a local minimum at a crossover transition, whereas it vanishes at the CP and on the associated spinodal lines. Indeed, lattice QCD shows that at vanishing baryon chemical potential, $\mu_B=0$, a minimum in $c_s$ occurs at temperature $T_0=156.5\pm1.5$ MeV \cite{Bazavov:2018mes} (see also \cite{Borsanyi:2020fev}), corresponding to a crossover transition between hadron gas and quark-gluon plasma (QGP). 

To date, a few attempts have been made to evaluate $c_s$ from heavy-ion collision data. In \cite{Gardim:2019xjs}, $c_s$ is estimated in ultrarelativistic collisions, where $\mu_B\approx n_B\approx0$, based on the proportionality of the entropy density, $s$, and the temperature, $T$, to the charged particle multiplicity and mean transverse momentum, respectively. The estimated value agrees with lattice QCD results. At finite $\mu_B$, the Landau model as well as hybrid hydrodynamics and hadronic transport simulations were used in \cite{Steinheimer:2012bp} to reproduce the widths of the negatively charged pion rapidity distribution. That study purports to locate a minimum in $c_s$ within the collision energy range $\sqrt{s_{NN}}=4\txt{-}9\ \txt{GeV}$.

In this Letter, we suggest a novel approach to exploring the behavior of $c_s$ by using cumulants of the baryon number distribution. The sensitivity of the cumulants to the EOS near the CP \cite{Asakawa:2009aj, Stephanov:2011pb}, which makes them central observables pursued in the Beam Energy Scan, follows directly from their sensitivity to  derivatives of the pressure with respect to $\mu_B$. The key observation in this Letter is that, besides the vicinity of the CP, cumulants provide rich information about the EOS at all points of the phase diagram, and in particular they allow a measurement of $c_s$ in matter created in heavy-ion collisions.

\section{Cumulants and the speed of sound}
\label{relationship_cumulants_cT2}

Cumulants of net baryon number $\kappa_j$ are defined as $\kappa_j=VT^{j-1}\left(d^jP/d\mu_B^j\right)_T$, where $V$ is the volume. Expressed in terms of derivatives with respect to $n_B$, the first three cumulants are given by
\begin{eqnarray}
&& \kappa_1  = V n_B ~, \label{cumulant_1} \hspace{8mm} \kappa_2 = \frac{VTn_B}{\left( \frac{dP}{dn_B} \right)_T}~, \label{cumulant_2} \non  \\
&& \kappa_3= \frac{VT^2n_B}{\left( \frac{dP}{dn_B} \right)_T^2} \left[  1 - \frac{n_B}{\left( \frac{dP}{dn_B} \right)_T} \left(\frac{d^2P}{dn_B^2}\right)_T \right] \label{cumulant_3} ~.
\end{eqnarray}
Importantly, cumulants are related to moments of the baryon number distribution. In particular, for $j\leq3$, $\kappa_j\equiv\big\langle\big(N_B-\big\langle N_B\big\rangle\big)^j\big\rangle$.

The definition of $c_s$ requires specifying which properties of the system are
considered constant during the propagation of the compression wave. One often uses the speed of
sound at constant entropy $S$ per net baryon number $N_B$, $c_{\sigma}^2\equiv\left(
dP/d\mathcal E\right)_{\sigma}$, where $\sigma=S/N_B$. Similarly, the speed of sound at constant temperature is $c_T^2\equiv\left(dP/d\mathcal E\right)_{T}$. These variants have specific regions of applicability. For example, the propagation of sound in air is governed by adiabatic
compression, so that using $c_{\sigma}^2$ is appropriate. On the other hand, when
there is a temperature reservoir (e.g.,\ in porous media) or when the cooling timescale is fast compared with the sound wave period (as is the case, e.g.,\ for an interstellar medium subject to radiative cooling), $c_T^2$ is applicable.

Explicitly, 
$c_{\sigma}^2$ and $c_T^2$
can be written as
\begin{eqnarray}
c_{\sigma}^2 =  \frac{  \Big( \frac{d P}{dn_B} \Big)_{T} \Big(\frac{d s}{d T} \Big)_{n_B}  +   \Big( \frac{ d P}{dT} \Big)_{n_B} \bigg[  \frac{s}{n_B} - \Big(\frac{d s}{d n_B}\Big)_T \bigg]  }{  \Big(\frac{sT}{n_B} + \mu_B  \Big) \Big( \frac{d s}{d T} \Big)_{n_B}   }  
\label{speed_isentropic} 
\end{eqnarray}
and
\begin{eqnarray}
c_T^2 = \frac{\Big( \frac{dP}{dn_B} \Big)_T}{ T  \Big(\frac{d s}{d n_B} \Big)_T   +  \mu_B  }~.
\label{speed_isothermal}
\end{eqnarray}
In the limit $T\to0$, the above expressions both lead to 
\begin{eqnarray}
c^2\Big|_{T=0} = \frac{1}{\mu_B} \bigg( \frac{dP}{dn_B} \bigg)_T ~. 
\label{cT2_approx}
\end{eqnarray}
Consequently, for $(\mu_B/T)\gg1$, the values of $c_{\sigma}^2$ and $c_T^2$ should largely coincide. Moreover, Eq.\ (\ref{speed_isothermal}) can be transformed to express $c_T^2$ as a function of the Eq.\ (\ref{cumulant_3}) cumulants,
\begin{eqnarray}
c_T^{2} = \left[\bigg(\parr{\log \kappa_1}{\log T}\bigg)_{\mu_B} + \frac{\mu_B}{T} \frac{\kappa_2}{\kappa_1} \right]^{-1}~.
\label{cT2_as_function_of_cumulants}
\end{eqnarray}
The first term in Eq.\ (\ref{cT2_as_function_of_cumulants}) is challenging to estimate from experimental data, however, it can be shown to be negligible for a degenerate Fermi gas, $(\mu_B/T)\gg1$, where it constitutes an order $\left(T/\mu_B\right)^2$ correction; then
\begin{eqnarray}
c_T^2 \approx \frac{T \kappa_1}{\mu_B \kappa_2}~.
\label{magic_equation_1}
\end{eqnarray}
We note that Eq.\ (\ref{magic_equation_1}) provides an upper limit to the value of $c_T^2$ as long as $(\partial \log \kappa_1/\partial \log T)_{\mu_B} > 0$.

Using Eq.\ (\ref{speed_isothermal}), one can also calculate the logarithmic derivative of $c_T^2$,
\begin{eqnarray}
\hspace{-1mm}\bigg(\frac{d \ln c_T^2}{d \ln n_B} \bigg)_T 
=  \frac{n_B \Big( \frac{d^2P}{dn_B^2} \Big)_T}{ \Big( \frac{dP}{dn_B} \Big)_T} - \frac{  \Big( \frac{dP}{dn_B} \Big)_T + Tn_B  \Big( \frac{d^2s}{dn_B^2} \Big)_T }{\mu_B + T  \Big( \frac{ds}{dn_B} \big)_T} ~.
\end{eqnarray}
It is again possible to rewrite the above equation in terms of the cumulants,
\begin{eqnarray}
\bigg(\frac{d \ln c_T^2}{d \ln n_B} \bigg)_T + c_T^2
= 1 - \frac{\kappa_3 \kappa_1}{\kappa_2^2}   - c_T^2 \bigg(\frac{d \ln (\kappa_2/T)}{d \ln T}\bigg)_{n_B} ~,
\end{eqnarray}
and neglecting the last term on the right-hand side yields
\begin{eqnarray}
\left(\frac{d \ln c_T^2}{d \ln n_B} \right)_T + c_T^2  \approx 1 - \frac{\kappa_3 \kappa_1}{\kappa_2^2} ~.
\label{magic_equation_2}
\end{eqnarray}
This approximation is again valid for $(\mu_B/T) \gg 1$, and the correction due to the neglected term is likewise of order $\left(T/\mu_B\right)^2$.

We note that in the opposite limit, $\mu_B \to 0$, Eq.\ (\ref{cT2_as_function_of_cumulants}) reveals a similarly simple form, $c_T^{2} = \left( d \ln \kappa_2/d \ln T \right)^{-1}_{\mu_B=0}$, suggesting that $c_T^2$ can be estimated in ultrarelativistic heavy-ion collisions, provided measurements of $\kappa_2$ are available at different temperatures. It might be possible to achieve this with data from a combination of centralities, energies, collision species, or rapidity ranges. In this work, however, we are interested in utilizing Eqs.\ (\ref{magic_equation_1}) and (\ref{magic_equation_2}) applied to collisions at medium and low energies.

\begin{figure*}[t]
	\includegraphics[width = 0.99\textwidth]{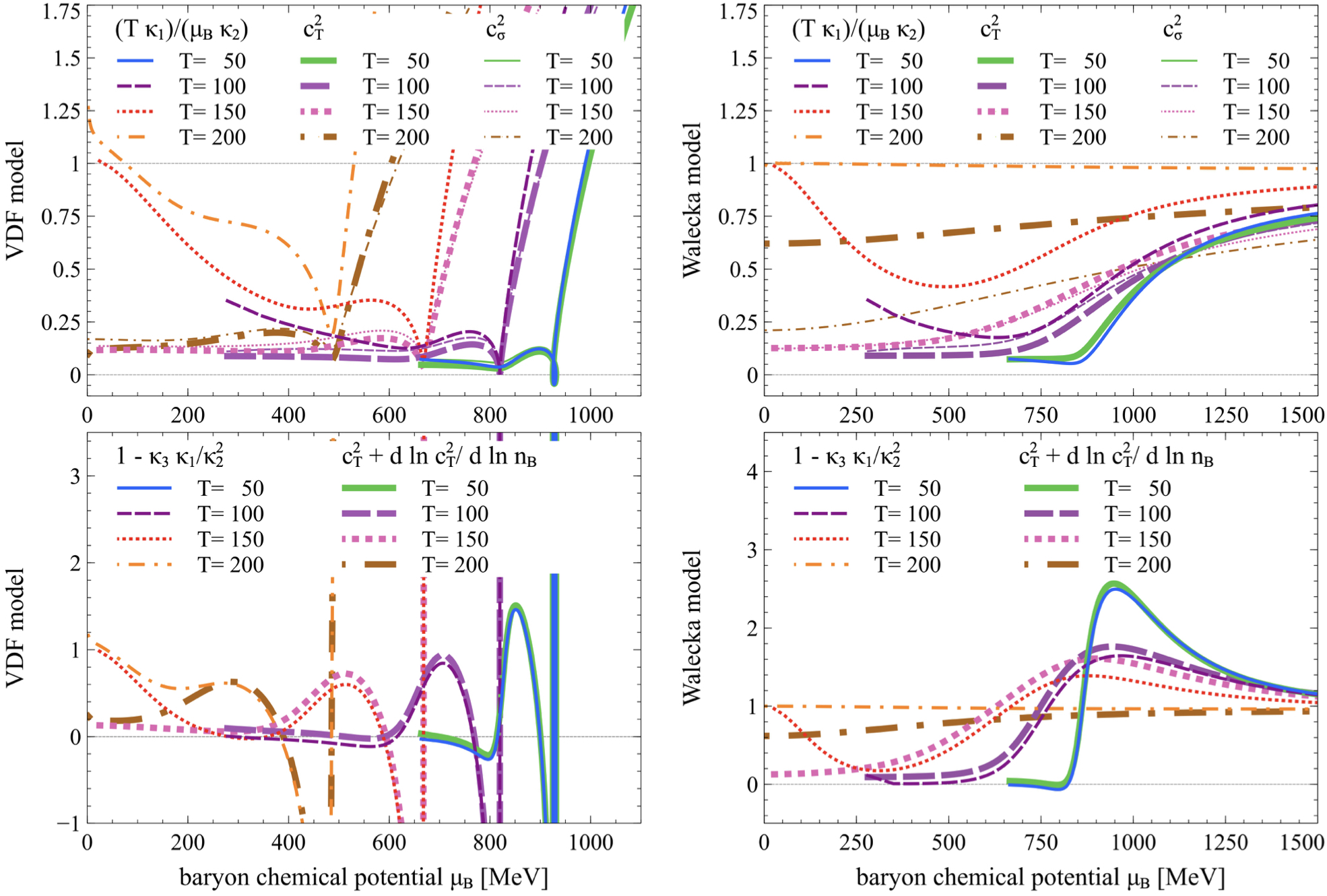}
	\caption{Model study of regions of applicability of Eqs.\ (\ref{magic_equation_1}) and (\ref{magic_equation_2}). The left (right) panels show results obtained in the VDF (Walecka) model. The upper and lower panels show quantities entering Eq.\ (\ref{magic_equation_1}) and Eq.\ (\ref{magic_equation_2}), respectively. Results at $T=50,100,150,200\ \txt{MeV}$ are given by blue and green solid lines, dark and light purple long-dashed lines, red and pink short-dashed lines, and orange and brown dash-dotted lines, respectively. For each $T$, the thickest lines correspond to the exact results and the medium-thick lines correspond to the approximations, given by the right-hand sides of Eqs.\ (\ref{magic_equation_1}) and (\ref{magic_equation_2}). Additionally, in upper panels the thinnest lines correspond to $c_{\sigma}^2$. Upper panels: for both models, Eq.\ (\ref{magic_equation_1}) is valid for $T\lesssim100\ \txt{MeV}$ and $\mu_B\gtrsim600\ \txt{MeV}$. Lower panels: for both models, Eq.\ (\ref{magic_equation_2}) is valid for $\mu_B\gtrsim200\ \txt{MeV}$; the exception is the Walecka model at $T=200\ \txt{MeV}$, where a phase transition to an almost massless gas of nucleons dramatically decreases the applicability of both Eqs.\ (\ref{magic_equation_1}) and (\ref{magic_equation_2}).
	}
	\label{tests_of_formulas}
\end{figure*}

\section{Validation}

We are interested in finding the limitations of the low-temperature approximation used to derive Eqs.\ (\ref{magic_equation_1}) and (\ref{magic_equation_2}), and for this we use effective models. Anticipating applying our formulas in regions of the phase diagram described by hadronic degrees of freedom, we choose two models of dense nuclear matter: the vector density functional (VDF) model with two phase transitions \cite{Sorensen:2020ygf} and the Walecka model \cite{Walecka:1974qa}. The VDF model utilizes interactions of the vector type, while the Walecka model employs both vector- and scalar-type interactions. Both models describe the nuclear liquid-gas phase transition, while the VDF model additionally describes a conjectured high-density, high-temperature phase transition modeling the QGP phase transition. In this work, the QGP-like phase transition is chosen to exhibit a CP at $T_c=100\ \txt{MeV}$ and  $n_c=3n_0$, with the $T=0$ boundaries of the spinodal region in the $T$-$n_B$ plane given by $n_{B, \txt{left spinodal} } (T=0) \equiv \eta_L=2.5 n_0$ and $n_{B, \txt{right spinodal}} (T=0) \equiv\eta_R=3.32 n_0$, where $n_0=0.160\ \txt{fm}^{-3}$; this choice is arbitrary and serves as a plausible example.

We plot both sides of Eqs.\ (\ref{magic_equation_1}) and (\ref{magic_equation_2}) as functions of $\mu_B$ at a series of
temperatures in Fig.\ \ref{tests_of_formulas}. We note that the explored temperature range reaches beyond the region of the phase diagram where hadronic models plausibly describe matter created in heavy-ion collisions; nevertheless, it is instructive to test our approximations in this regime.

We use natural units in which the speed of light in vacuum is $c=1$. We note that in the VDF model, $c_s$ quickly becomes acausal for $\mu_B$ above the QGP-like phase transition. It is an expected behavior in models using interactions dependent on high powers of $n_B$ \cite{Zeldovich:1962emp}, and while not ideal, it does not affect the current analysis.

In all panels in Fig.\ \ref{tests_of_formulas}, the exact model calculations show expected
features as functions of $\mu_B$. In the upper left panel, showing both $c_{T}^2$ and $c_{\sigma}^2$ in the VDF model, at small $\mu_B$ we see a softening of the EOS due to the influence of the nuclear CP, followed by an increase at densities of the order of $n_0$, then a dive in $c_s^2$ caused by the QGP-like phase transition, and finally a steep rise for high values of $\mu_B$. In the upper right panel, showing $c_T^2$ and $c_{\sigma}^2$ in the Walecka model, we similarly observe a soft EOS at small $\mu_B$, while the value of $c_s^2$ goes asymptotically to 1 for large $\mu_B$. Additionally, for $T=200\ \txt{MeV}$, the Walecka model shows effects due to a phase transition in the nucleon-antinucleon plasma, occurring around $T\approx190\ \txt{MeV}$ and $n_B=0$; above this transition, the model describes an almost noninteracting gas of nearly massless nucleons \cite{Theis:1984qc}. The behavior of the curves in the lower panels, showing $c_T^2+\big(\frac{d\ln c_T^2}{d\ln n_B}\big)_T$, can be directly traced to the behavior of the curves in the upper panels. In particular, for the VDF model we observe strong divergences due to the softening of the EOS in the QGP-like phase transition region.

Comparing the exact results to the approximations, we see that, while Eq.\ (\ref{magic_equation_1}) is valid for $T\lesssim100\ \txt{MeV}$ and $\mu_B\gtrsim600\ \txt{MeV}$, it behaves poorly, both qualitatively and quantitatively, for $T$ and $\mu_B$ corresponding to regions of the phase diagram probed by moderately to highly energetic heavy-ion collisions (upper panels). On the other hand, the approximation introduced in Eq.\ (\ref{magic_equation_2}) is qualitatively valid for most of the probed $T$ and $\mu_B$, with the exception of regions characterized by $\mu_B\lesssim200\ \txt{MeV}$ (lower panels).

\begin{figure}[t]
	\includegraphics[width = 0.99\columnwidth]{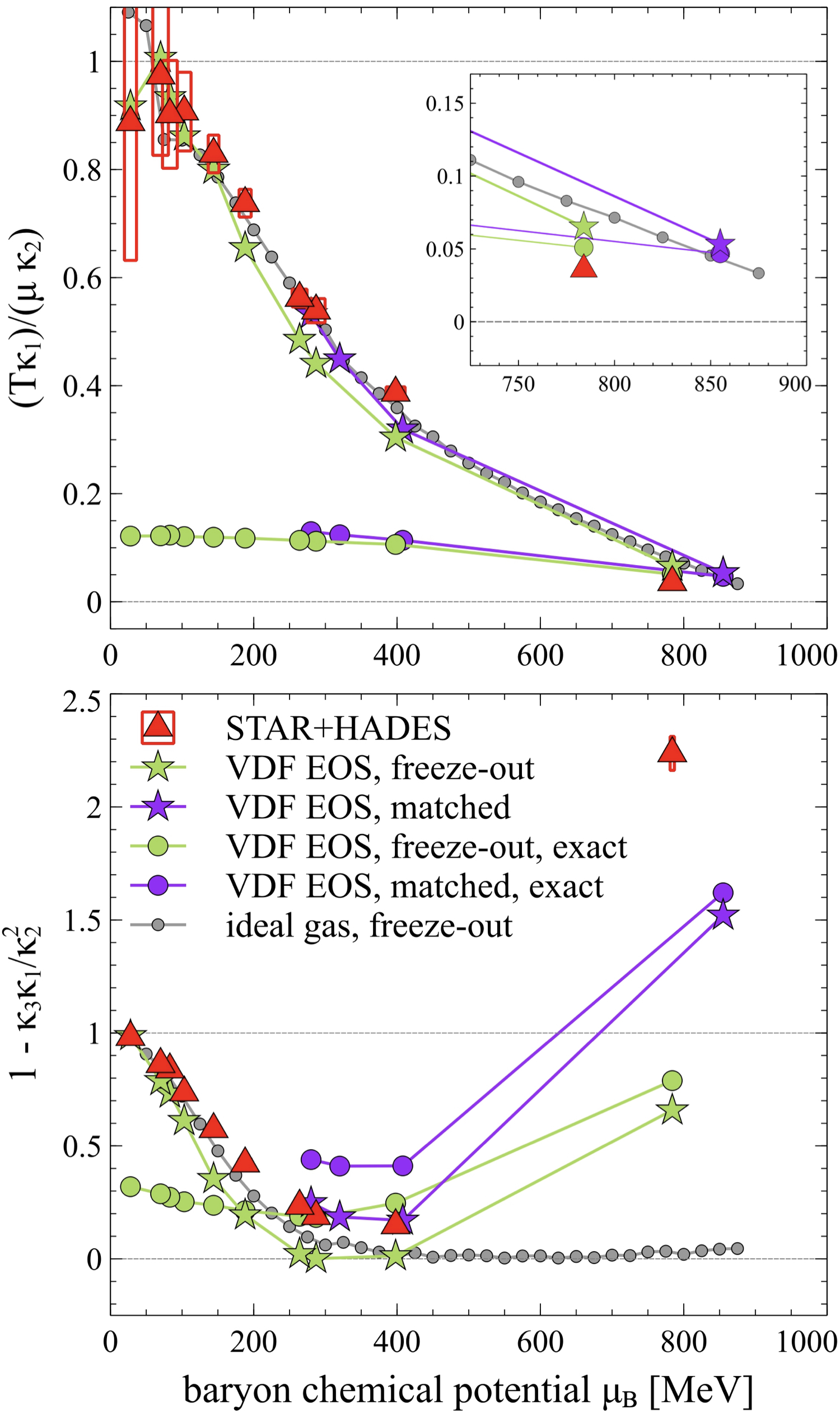}\\
	\caption{Comparison of the right-hand sides of Eq.\ (\ref{magic_equation_1}) (upper panel) and Eq.\ (\ref{magic_equation_2}) (lower panel) for experimental data (red triangles), ideal gas at the freeze-out (small gray circles), the VDF model at the freeze-out (light green stars), and the VDF model at a set of points chosen to reproduce the data (dark purple stars); exact results, that is, the left-hand sides of Eqs.\ (\ref{magic_equation_1}) and (\ref{magic_equation_2}), are shown for the two cases considered in the VDF model (green and purple circles). The data points for the matched VDF results (shown only for collisions at low energies, where using the model is justified) are chosen to reproduce experimental values of $1-\kappa_3\kappa_1/\kappa_2^2$ (see Fig.\ \ref{diagram}). We note that at $\sqrt{s}=2.4\ \txt{GeV}$, matching the value of $1-\kappa_3 \kappa_1/\kappa_2^2$ exactly is possible, but would place the matched point close to the nuclear liquid-gas CP, which we find unlikely.  
	}
	\label{STAR_HADES_plots}
\end{figure}

\begin{figure}[t]
	\includegraphics[width = 0.99\columnwidth]{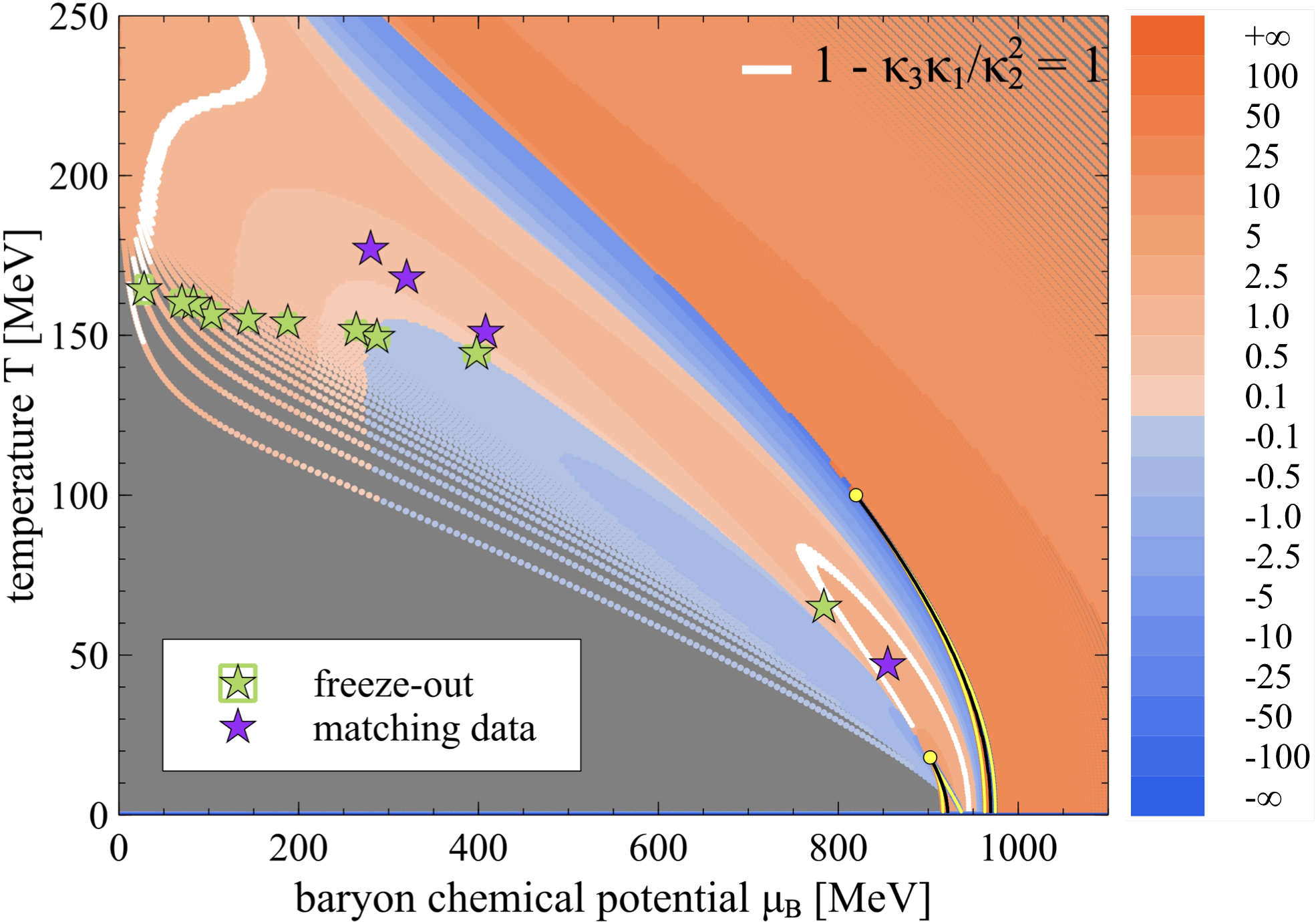}
	\caption{Contour plot of $1-\kappa_3\kappa_1/\kappa_2^2$ in the VDF model. Yellow and black lines correspond to the spinodal and coexistence lines, respectively; white contours signify regions where $1-\kappa_3\kappa_1/\kappa_2^2=1\pm0.03$. Light green stars denote experimentally measured freeze-out parameters $(T_{\txt{fo}},\mu_{\txt{fo}})$, while dark purple stars denote points where $1-\kappa_3\kappa_1/\kappa_2^2$, taken along lines informed by average phase diagram trajectories for STAR collision energies \cite{Shen:2020jwv}, matches the experimentally measured values for a given collision energy. The softening of the EOS, leading to negative values of $\big(\frac{d\ln c_T^2}{d\ln n_B}\big)_T$, occurs in two regions of the phase diagram, corresponding to the ordinary nuclear matter phase transition and to the conjectured QGP-like phase transition. 
	}
	\label{diagram}
\end{figure}

\section{Experimental data and interpretation}

We consider cumulants of the net proton number and chemical freeze-out parameters, $(T_{\txt{fo}},\mu_{\txt{fo}})$, in collisions at 0-5\% centrality,  determined by the solenoidal tracker at RHIC (STAR) \cite{Abdallah:2021fzj} and high acceptance dielectron spectrometer (HADES) \cite{Adamczewski-Musch:2020slf, HADES_MLorentz_talk} experiments, and we use them to plot Eqs.\ (\ref{magic_equation_1}) and (\ref{magic_equation_2}) (red triangles, upper and lower panel in Fig.\ \ref{STAR_HADES_plots}, respectively) against $\mu_B$.

Based on the previous section, we trust the results presented in the upper panel of Fig.\ \ref{STAR_HADES_plots}, approximating $c_T^2$, only for the lowest collision energy, $\sqrt{s}=2.4\ \txt{GeV}$ from the HADES experiment. Here, $c_T^2$ as obtained from Eq.\ (\ref{magic_equation_1}) is small: less than half of the ideal gas value. At the same time, the value of $1-\kappa_3\kappa_1/\kappa_2^2$ (shown in the lower panel in Fig.\ \ref{STAR_HADES_plots}), which we assume is dominated by $\big(\frac{d\ln c_T^2}{d\ln n_B}\big)_T$, drops with decreasing collision energy to reach a minimum at the lowest STAR point, $\sqrt{s}=7.7\ \txt{GeV}$, and then steeply rises for the HADES point. This could mean that in $\sqrt{s} = 7.7\ \txt{GeV}$ collisions, $c_T^2$ is approximately constant as a function of $n_B$, while in $\sqrt{s} = 2.4\ \txt{GeV}$ collisions the matter is characterized by a small $c^2_T$ which nevertheless has a large slope as a function of $n_B$.

To further understand this behavior, we study the dependence of $1-\kappa_3\kappa_1/\kappa_2^2$ on $\mu_{B}$ and $T$ within the VDF model, shown in Fig.\ \ref{diagram}. Comparing model results to experiment requires choosing at which $T$ and $\mu_B$ to take values of $1-\kappa_3 \kappa_1/\kappa_2^2$. A natural choice is to use values at $(T_{\txt{fo}},\mu_{\txt{fo}})$ (light green stars), but as shown in the lower panel of Fig.\ \ref{STAR_HADES_plots}, these values do not lead to an agreement with experimental data, with the biggest discrepancy for the HADES point. (We note that $(T_{\txt{fo}},\mu_{\txt{fo}})$ are established with hadron interactions neglected, and the degree to which this affects our results may vary across the phase diagram.) However, critical fluctuations exhibit a large relaxation time \cite{Berdnikov:1999ph,Stephanov:2017ghc,Du:2020bxp}, and their measured values could be affected by stages of the collision preceding the freeze-out. With this insight and taking guidance from average phase diagram trajectories of hybrid simulations of heavy-ion collisions \cite{Shen:2020jwv}, we consider $1-\kappa_3\kappa_1/\kappa_2^2$ at slightly earlier stages of the evolution. In this way we obtain values of $1-\kappa_3\kappa_1/\kappa_2^2$ (dark purple stars) that reproduce the experimental results for a given collision energy (lower panel in Fig.\ \ref{STAR_HADES_plots}); here the exception is the HADES point, for which we prioritized choosing a point in a reasonable vicinity of the measured freeze-out over obtaining a value equal to the experimental data. Comparing to the exact model results for $c_T^2$ and $c_T^2 + \big(\frac{d\ln c_T^2}{d\ln n_B}\big)_T$, also displayed in Fig.\ \ref{STAR_HADES_plots}, as well as to the upper left panel of Fig.\ \ref{tests_of_formulas}, we can confirm that at the point reproducing the experimental value of $1-\kappa_3\kappa_1/\kappa_2^2$ for the lowest STAR energy, $c_T^2$ is nearly constant as a function of $\mu_B$ (thick short-dashed line, $T=150\ \txt{MeV}$, at $\mu_B\approx410\ \txt{MeV}$ in Fig.\ \ref{tests_of_formulas}), while at the point reproducing the result for the HADES energy, $c_T^2$ increases sharply with $\mu_B$ (thick solid line, $T=50\ \txt{MeV}$, at $\mu_B\approx850\ \txt{MeV}$ in Fig.\ \ref{tests_of_formulas}).

Naturally, the choice of $T$ and $\mu_B$ at which we compare model calculations with STAR and HADES cumulant data is driven by the wish to match the experimental results, and it serves mainly to show that baryon number cumulants measured in heavy-ion collisions can be connected to the speed of sound in hot and dense nuclear matter. Whether values of higher order cumulants are indeed significantly affected by stages of the evolution preceding the freeze-out needs to be further investigated (for recent developments, see \cite{Nahrgang:2018afz,Jiang:2017sni}). Moreover, while experiments measure proton number cumulants, the VDF model provides baryon number cumulants, putting more strain on our interpretation. Baryon number conservation should likewise be important \cite{Bzdak:2012an,Vovchenko:2020gne}. Finally, our model results may not be applicable in regions of the phase diagram where quarks and gluons become increasingly relevant.

Nonetheless, hadronic models are well-justified for describing low-energy collisions whose evolution is dominated by the hadronic stage. The comparison between the experimental data and the VDF model suggests that collisions at the lowest STAR and HADES energies may be probing regions of the phase diagram where the cumulants of the baryon number tell us more about hadronic physics than the QCD CP. In particular, the change in the sign of $\kappa_3$, predicted to take place in the vicinity of a critical point \cite{Asakawa:2009aj} and apparent in the HADES data (see lower panel of Fig.\ \ref{STAR_HADES_plots}), may mark the region of the phase diagram affected by the nuclear liquid-gas phase transition. If this is true, it may be worthwhile to study the cumulants at even lower collision energies, starting from $0.1\ \txt{GeV}$ projectile kinetic energy, and obtain the speed of sound around the nuclear liquid-gas CP. Conversely, at higher energies it could be possible to use collisions at different centralities and different rapidity windows to estimate the neglected terms in Eqs.\ (\ref{magic_equation_1}) and (\ref{magic_equation_2}), and obtain a stronger estimate for the speed of sound in the respective regions of the phase diagram.

\section{Summary and conclusions}

In this work, we use cumulants of the baryon number distribution to estimate the isothermal speed of sound squared and its logarithmic derivative with respect to the baryon number density. This result provides a new method for obtaining information about fundamental properties of nuclear matter studied in heavy-ion collisions, with consequences for both the search for the QCD CP and neutron star studies. While the approximations and the model comparison we considered apply to experiments at low energies, the approach itself can be used at any collision energy provided that measurements of cumulants of baryon number distribution as well as their temperature dependence are available. Further studies of effects due to dynamics, in particular using state-of-the-art simulations, will be absolutely essential in determining the extent to which the proposed method provides a reliable extraction of sound velocities and their derivatives.

{\hspace{10mm}}

\section{Acknowledgements}

A.S.\ thanks Chun Shen for providing average phase diagram trajectories of hybrid heavy-ion collision simulations at RHIC Beam Energy Scan energies. 

A.S.\ and V.K.\ received support through the U.S. Department of Energy, 
Office of Science, Office of Nuclear Physics, under contract no.\ 
DE-AC02-05CH11231231 and received support within the framework of the
Beam Energy Scan Theory (BEST) Topical Collaboration. D.O.\ and L.M.\ were supported by the U.S.\ DOE under Grant No.\ DE-FG02-00ER4113.

\bibliography{inspire_citations_AS,non_inspire_AS}

\end{document}